\documentclass[times,10pt,twocolumn]{article}
\usepackage{sty/latex8}
\usepackage{sty/titling}
\usepackage[normalem]{ulem}
\usepackage[top=0.75in, bottom=0.75in, left=0.70in, right=0.70in]{geometry}
\usepackage{mdwlist}
\usepackage{url}
\usepackage{xspace}
\usepackage[compact]{titlesec}
\titlespacing{\subsection}{0pt}{4pt}{2pt}
\usepackage{graphicx}
\usepackage{wrapfig}
\usepackage{stfloats}
\usepackage[labelfont={small,bf}, textfont={small,bf},labelsep=period]{caption}
\usepackage{subcaption}
\usepackage{booktabs}
\usepackage{multirow}
\usepackage[usenames, dvipsnames]{color}
\usepackage{times}
\usepackage[scaled=0.90]{helvet}
\usepackage{bbding}
\usepackage{pifont}

\usepackage{amsmath}
\usepackage{amssymb}
\usepackage{stmaryrd}
\usepackage{indentfirst}
\usepackage{footnote}
\newcommand{\mycolor}[1]{\textcolor{black}{#1}}

\newcommand{\ignore}[1]{}   

\newcommand\trc{$t_{\mathit{RC}}$\xspace}
\newcommand\trcd{$t_{\mathit{RCD}}$\xspace}

\usepackage{setspace}
\usepackage{sty/bibspacing}
\usepackage{fancyhdr}

\fancypagestyle{plain}{
	\fancyhf{}
	\chead{SAFARI Technical Report No. 2016-001 (January 26, 2016)}
	\cfoot{\thepage}
	
}

\begin{document}
\date{}
\pagestyle{plain}

\pretitle{\begin{center}\vspace{-0.5in} This is a summary of the original paper,
entitled ``Tiered-Latency DRAM: A Low Latency and Low Cost DRAM Architecture''
which appears in HPCA 2013~\cite{tldram}.\\}

\title{\Large\bf \vskip 0.15in
Tiered-Latency DRAM (TL-DRAM)\\
\vskip -0.15in\vspace{-0.0in}}

\author{
\hspace{0.1in} \vspace{-0.15in}
Donghyuk Lee 					\and \hspace{-0.25in}
Yoongu Kim 						\and \hspace{-0.25in}
Vivek Seshadri 				\and \hspace{-0.25in}
Jamie Liu 						\and \hspace{-0.25in}
Lavanya Subramanian 	\and \hspace{-0.25in}
Onur Mutlu 						\and \hspace{0.1in}
Carnegie Mellon University \vspace*{-2.5in} \\
}
\affiliation{}

\maketitle
\setstretch{0.97}

\begin{abstract}

This paper summarizes the idea of Tiered-Latency DRAM, which was
published in HPCA 2013~\cite{tldram}. The key goal of TL-DRAM is to provide
{\em low DRAM latency at low cost}, a critical problem in modern memory
systems~\cite{mutlu-superfri2015}. To this end, TL-DRAM introduces {\em
heterogeneity} into the design of a DRAM subarray by segmenting the bitlines,
thereby creating a low-latency, low-energy, low-capacity portion in the
subarray (called the {\em near segment}), which is close to the sense
amplifiers, and a high-latency, high-energy, high-capacity portion, which is
farther away from the sense amplifiers. Thus, DRAM becomes heterogeneous with a
small portion having lower latency and a large portion having higher latency.
Various techniques can be employed to take advantage of the low-latency near
segment and this new heterogeneous DRAM substrate, including hardware-based
caching and software based caching and memory allocation of frequently used
data in the near segment. Evaluations with simple such techniques show
significant performance and energy-efficiency benefits~\cite{tldram}.

\end{abstract}

\section{Summary}

\subsection{The Problem: High DRAM Latency} \label{sec:problem}

Primarily due to its low cost-per-bit, DRAM has long been the choice substrate
for architecting main memory subsystems. In fact, DRAM's cost-per-bit has been
decreasing at a rapid rate as DRAM process technology scales to integrate ever
more DRAM cells into the same die area. As a result, each successive generation
of DRAM has enabled increasingly large-capacity main memory subsystems at low
cost.

In stark contrast to the continued scaling of cost-per-bit, the {\em latency}
of DRAM has remained almost constant. During the same 11-year interval in which
DRAM's cost-per-bit decreased by a factor of 16, DRAM latency (as measured by
the \trcd and \trc timing constraints) decreased by only 30.5\% and
26.3\%~\cite{future_cpu, samsung_roadmap}, as shown in Figure 1 of our
paper~\cite{tldram}. From the perspective of the processor, an access to DRAM
takes hundreds of cycles --- time during which the processor may be stalled,
waiting for DRAM. Such wasted time leads to large performance degradations.

\subsection{Key Observations and Our Goal} \label{sec:observation}

{\bf Bitline: Dominant Source of Latency.} In DRAM, each bit is represented as
electrical charge in a capacitor-based {\em cell}. The small size of this
capacitor necessitates the use of an auxiliary structure, called a {\em
sense-amplifier}, to detect the small amount of charge held by the cell and
amplify it to a full digital logic value. But, a sense-amplifier is
approximately one hundred times larger than a cell~\cite{rambus-power}. To
amortize their large size, each sense-amplifier is connected to many DRAM cells
through a wire called a {\em bitline}.

Every bitline has an associated parasitic capacitance whose value is
proportional to the length of the bitline. Unfortunately, such parasitic
capacitance slows down DRAM operation for two reasons. First, it increases the
latency of the sense-amplifiers. When the parasitic capacitance is large, a
cell cannot quickly create a voltage perturbation on the bitline that could be
easily detected by the sense-amplifier. Second, it increases the latency of
charging and precharging the bitlines. Although the cell and the bitline must
be restored to their quiescent voltages during and after an access to a cell,
such a procedure takes much longer when the parasitic capacitance is large. Due
to the above reasons and a detailed latency break-down (refer to our HPCA-19
paper~\cite{tldram}), we conclude that long bitlines are the dominant source of
DRAM latency~\cite{jedec-ddr, dram_latency, mutlu-imw13, mutlu-book15}.

{\bf Latency vs.~Cost Trade-Off.} The bitline length is a key design parameter
that exposes the important trade-off between latency and die-size (cost). Short
bitlines (few cells per bitline) constitute a small electrical load (parasitic
capacitance), which leads to low latency. However, they require more
sense-amplifiers for a given DRAM capacity
(Figure~\ref{fig:intro_specialized_dram}), which leads to a large die-size. In
contrast, long bitlines have high latency and a small die-size
(Figure~\ref{fig:intro_commodity_dram}). As a result, neither of these two
approaches can optimize for both latency and cost-per-bit.

\begin{figure}[ht]
  \centering
  \begin{subfigure}[b]{1in}
    \centering
    \captionsetup{font=footnotesize}
    \includegraphics[width=0.9in]{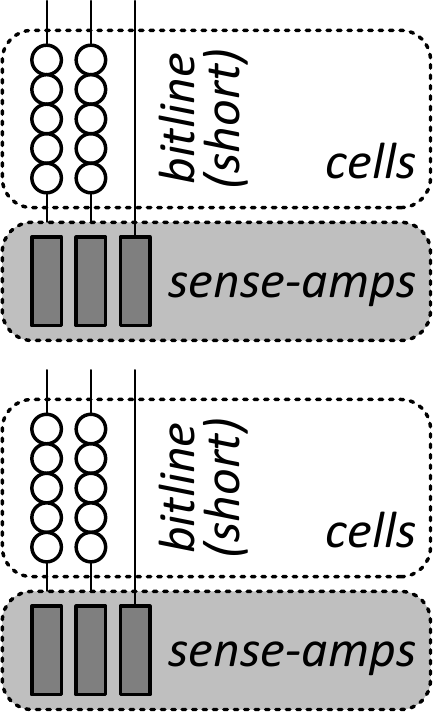}
    \subcaption{Latency Opt.}
    \label{fig:intro_specialized_dram}
  \end{subfigure}
  \begin{subfigure}[b]{1.1in}
    \centering
    \includegraphics[width=0.9in]{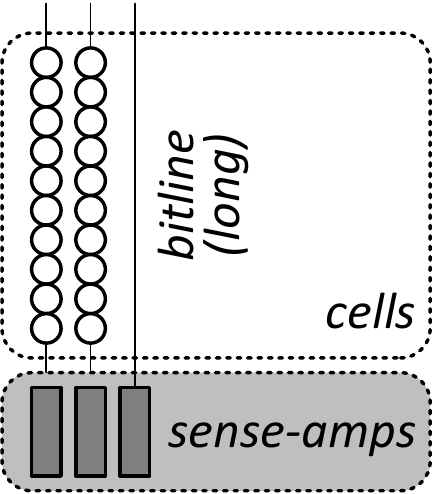}
    \subcaption{Cost Optimized}
    \label{fig:intro_commodity_dram}
  \end{subfigure}
  \begin{subfigure}[b]{1in}
    \centering
    \includegraphics[width=0.9in]{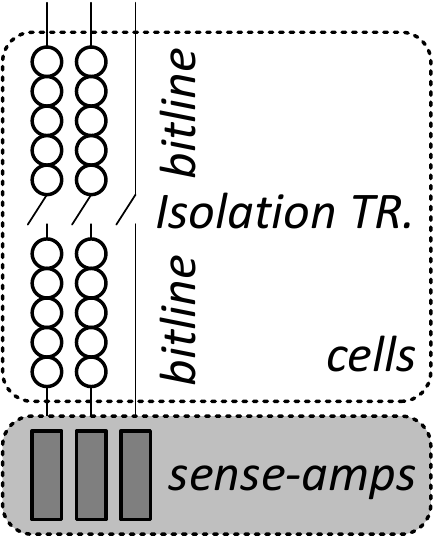}
    \subcaption{Our Proposal}
    \label{fig:intro_tldram}
  \end{subfigure}
  \caption{DRAM: Latency vs. Cost Optimized, Our Proposal}
  \label{fig:intro_commodity_specialized}
\end{figure}

Figure~\ref{fig:cell-per-bitline-trade-off} shows the trade-off between DRAM
latency and die-size by plotting the latency (\trcd and \trc) and the die-size
for different values of cells-per-bitline. Existing DRAM architectures are
either optimized for die-size (commodity DDR3~\cite{samsung_spec, ddr3-4gb})
and are thus low cost but high latency, or optimized for latency
(RLDRAM~\cite{rldram}, FCRAM~\cite{fcram}) and are thus low latency but high
cost.

\begin{figure}[ht]
	\vspace{0.1in}
  \centering
  \includegraphics[width=0.9\linewidth]{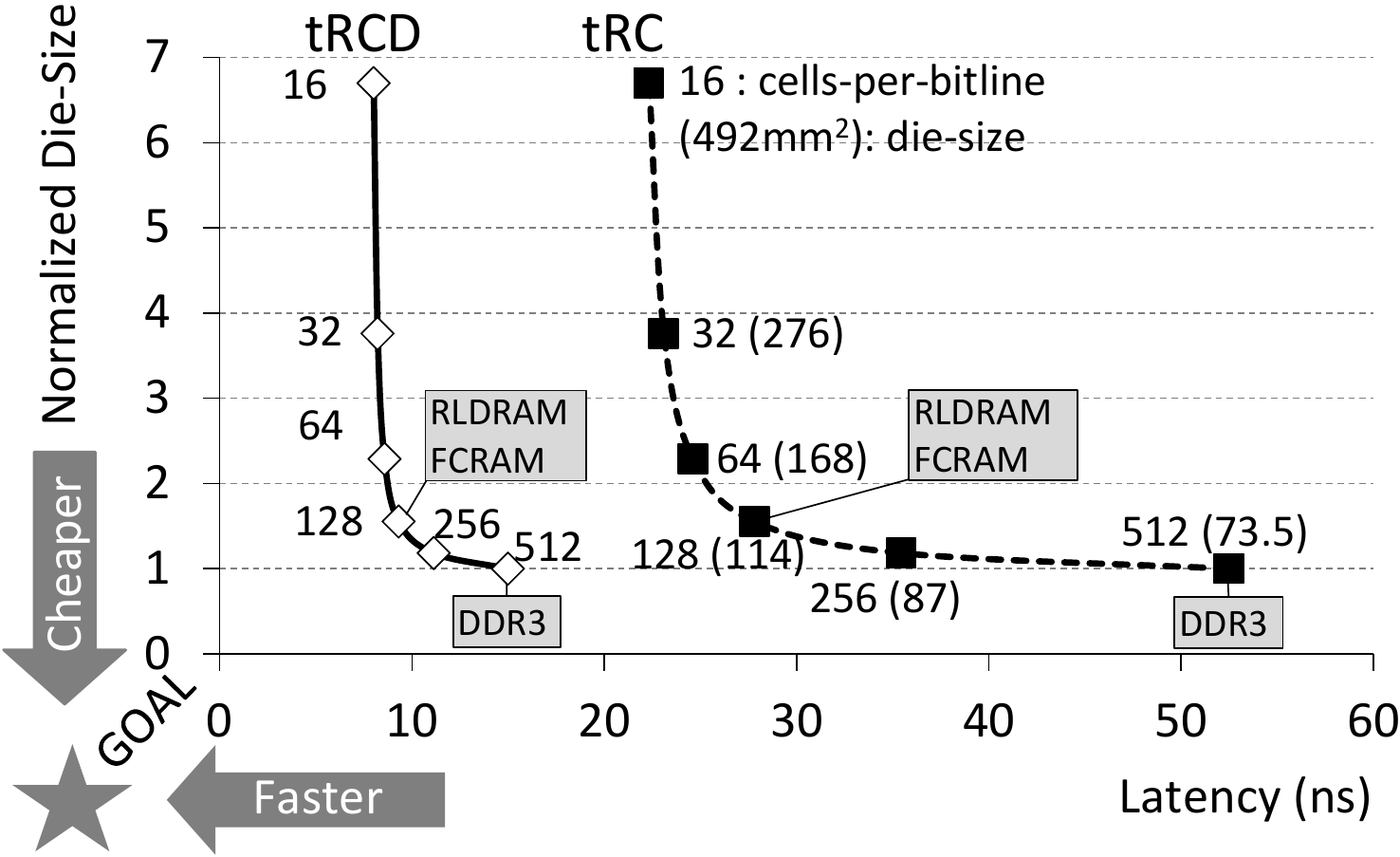} \\
  \caption{Bitline Length: Latency vs. Die-Size}
  \label{fig:cell-per-bitline-trade-off}
\end{figure}

{\bf The goal} of our paper~\cite{tldram} is to design a new DRAM architecture
to approximate the best of both worlds (i.e., low latency and low cost), based
on the key observation that that long bitlines are the dominant source of DRAM
latency.

\subsection{Tiered-Latency DRAM} \label{sec:tldram}

To achieve the latency advantage of short bitlines and the cost advantage of
long bitlines, we propose the {\em Tiered-Latency DRAM} (TL-DRAM) architecture,
which is shown in Figure~\ref{fig:intro_tldram} and \ref{fig:substrate_tld}. The
key idea of TL-DRAM is to divide the long bitline into two shorter segments
using an {\em isolation transistor}: the {\em near segment} (connected directly
to the sense-amplifier) and the {\em far segment} (connected through the
isolation transistor).

\begin{figure}[ht]
  \begin{subfigure}[b]{0.9in}
    \centering
    \includegraphics[height=1.2in]{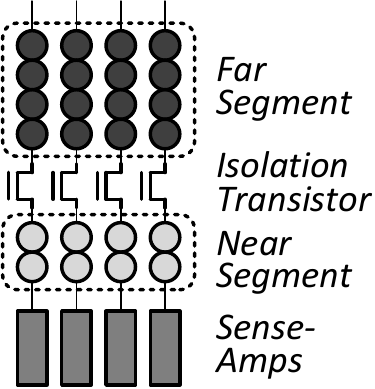}
    \subcaption{Organization}
    \label{fig:substrate_tld}
  \end{subfigure}\qquad
  \begin{subfigure}[b]{1.1in}
    \centering
    \includegraphics[height=1.2in]{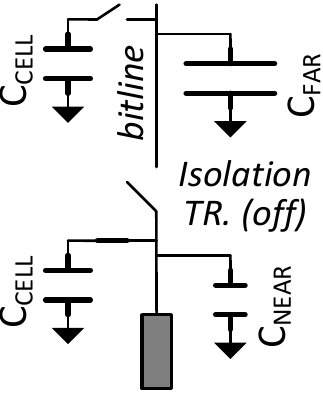}
    \caption{Near Seg. Access}
    \label{fig:substrate_tld_near}
  \end{subfigure}
  \begin{subfigure}[b]{1.1in}
    \centering
    \includegraphics[height=1.2in]{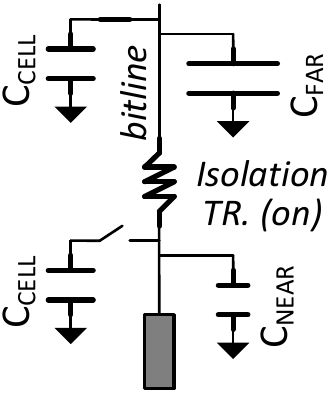}
    \caption{Far Seg. Access}
    \label{fig:substrate_tld_far}
  \end{subfigure}
  \caption{TL-DRAM: Near vs. Far Segments}
  \label{fig:substrate_tldram}
\end{figure}

The primary role of the isolation transistor is to electrically decouple the
two segments from each other. This changes the effective bitline length (and
also the effective bitline capacitance) as seen by the cell and
sense-amplifier. Correspondingly, the latency to access a cell is also changed
--- albeit differently depending on whether the cell is in the near or the far
segment.

When accessing a cell in the near segment, the isolation transistor is turned
off, disconnecting the far segment (Figure~\ref{fig:substrate_tld_near}). Since
the cell and the sense-amplifier see only the reduced bitline capacitance of
the shortened near segment, they can drive the bitline voltage more easily.  As
a result, the bitline voltage is restored more quickly, so that the latency
(\trc) for the near segment is significantly reduced. On the other hand, when
accessing a cell in the far segment, the isolation transistor is turned on to
connect the entire length of the bitline to the sense-amplifier. In this case,
the isolation transistor acts like a resistor inserted between the two segments
(Figure~\ref{fig:substrate_tld_far}) and limits how quickly charge flows to the
far segment. Because the far segment capacitance is charged more slowly, it
takes longer for the far segment voltage to be restored, so that the latency
(\trc) is increased for cells in the far segment.

{\bf Latency, Power, and Die-Area.} Table~\ref{tbl:latency_comparison}
summarizes the latency, power, and die-area characteristics of TL-DRAM to other
DRAMs, estimated using circuit-level SPICE simulation~\cite{ibm_55nm} and
power/area models from Rambus~\cite{rambus-power}. Compared to commodity DRAM
(long bitlines) which incurs high latency (\trc) for all cells, TL-DRAM offers
significantly reduced latency (\trc) for cells in the near segment, while
increasing the latency for cells in the far segment due to the additional
resistance of the isolation transistor. In DRAM, a large fraction of the power
is consumed by the bitlines. Since the near segment in TL-DRAM has a lower
capacitance, it also consumes less power. On the other hand, accessing the far
segment requires toggling the isolation transistors, leading to increased power
consumption. Mainly due to additional isolation transistors, TL-DRAM increases
die-area by 3\%. Our paper includes detailed circuit-level analyses of TL-DRAM
(Section 4 of~\cite{tldram}).

\begin{table}[ht]
\begin{footnotesize}
\setlength{\tabcolsep}{4.5pt}

\centering
\begin{tabular}{cccccc}

\toprule
& & Short Bitline & Long Bitline & \multicolumn{2}{c}{Segmented Bitline} \\
& &
(Fig~\ref{fig:intro_specialized_dram}) &
(Fig~\ref{fig:intro_commodity_dram}) &
\multicolumn{2}{c}{(Fig~\ref{fig:intro_tldram})} \\
\cmidrule{3-6}
& & Unsegmented & Unsegmented& Near & Far  \\
\cmidrule{1-6}

\multicolumn{2}{c}{Length (Cells)} & 32 & 512 & 32 & 480  \\

\cmidrule{1-6}

\multicolumn{2}{c}{Latency} & {\bf Low}  & High   & {\bf Low}  & Higher \\
\multicolumn{2}{c}{(\trc)}  & (23.1ns) & (52.5ns) & (23.1ns) & (65.8ns) \\

\cmidrule{1-6}

\multicolumn{2}{c}{Normalized}     & {\bf Low} & High   & {\bf Low} & Higher  \\
\multicolumn{2}{c}{Power Consump.} & (0.51) & (1.00) & (0.51) & (1.49)  \\

\cmidrule{1-6}

\multicolumn{2}{c}{Normalized}      & High    & {\bf Lower} & \multicolumn{2}{c}{{\bf Low}}      \\
\multicolumn{2}{c}{Die-Size (Cost)} & (3.76)  & (1.00) & \multicolumn{2}{c}{(1.03)}   \\

\bottomrule
\end{tabular}
\end{footnotesize}
\caption{Latency, Power, and Die-Area Comparison} \label{tbl:latency_comparison}
\end{table}

\subsection{Leveraging TL-DRAM} \label{sec:mechanism}

TL-DRAM enables the design of many new memory management policies that exploit
the asymmetric latency characteristics of the near and the far segments. Our
HPCA-19 paper (in Section 5) describes four ways of taking advantage of
TL-DRAM. Here, we describe two approaches in particular.

In the first approach, the memory controller uses the near segment as a {\em
hardware-managed cache} for the far segment. In our HPCA-19
paper~\cite{tldram}, we discuss three policies for managing the near segment
cache. (The three policies differ in deciding when a row in the far segment is
cached into the near segment and when it is evicted.) In addition, we propose a
new data transfer mechanism ({\em Inter-Segment Data Transfer}) that
efficiently migrates data between the segments by taking advantage of the fact
that the bitline is a bus connected to the cells in both segments. By using
this technique, the data from the source row can be transferred to the
destination row over the bitlines at very low latency (additional 4ns over
\trc). Furthermore, this Inter-Segment Data Transfer happens exclusively within
DRAM bank without utilizing the DRAM channel, allowing concurrent accesses to
other banks.

In the second approach, the near segment capacity is exposed to the OS,
enabling the OS to use the full DRAM capacity. We propose two concrete
mechanisms, one where the memory controller uses an additional layer of
indirection to map frequently accessed pages to the near segment, and another
where the OS uses static/dynamic profiling to directly map frequently accessed
pages to the near segment. In both approaches, the accesses to pages that are
mapped to the near segment are served faster and with lower power than in
conventional DRAM, resulting in improved system performance and energy
efficiency.

\subsection{Results: Performance and Power}

Our HPCA-19 paper~\cite{tldram} provides extensive detail about both of the
above approaches. But, due to space constraints, we present the evaluation
results of only the first approach, in which the near segment is used as
hardware-managed cache managed under our best policy ({\em Benefit-Based
Caching}) to show the advantage of our TL-DRAM substrate.

{\bf Performance \& Power Analysis.} Figure~\ref{fig:result_cores} shows the
average performance improvement and power-efficiency of our proposed mechanism
over the baseline with conventional DRAM, on 1-, 2- and 4-core systems. As
described in Section~\ref{sec:tldram}, access latency and power consumption are
significantly lower for near segment accesses, but higher for far segment
accesses, compared to accesses in a conventional DRAM. We observe that a large
fraction (over 90\% on average) of requests hit in the rows cached in the near
segment, thereby accessing the near segment with low latency and low power
consumption. As a result, TL-DRAM achieves significant performance improvement
by 12.8\%/12.3\%/11.0\% and power savings by 23.6\%/26.4\%/28.6\% in
1-/2-/4-core systems, respectively.

\begin{figure}[ht]
  \begin{subfigure}[b]{0.48\linewidth}
    \centering
    \includegraphics[width=\linewidth]{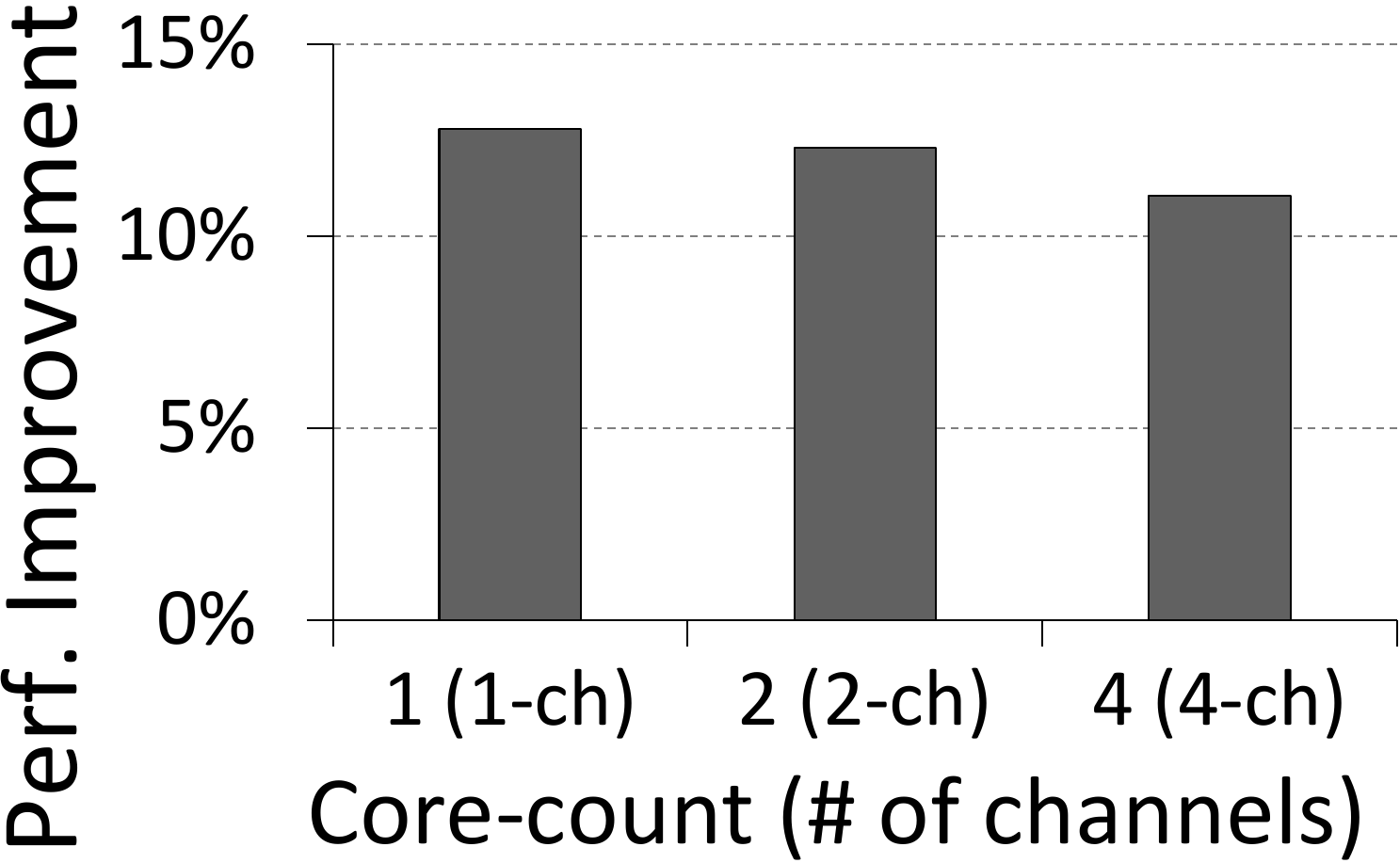}
    \subcaption{IPC Improvement}
    \label{fig:result_ipc}
  \end{subfigure}
  \begin{subfigure}[b]{0.48\linewidth}
    \includegraphics[width=\linewidth]{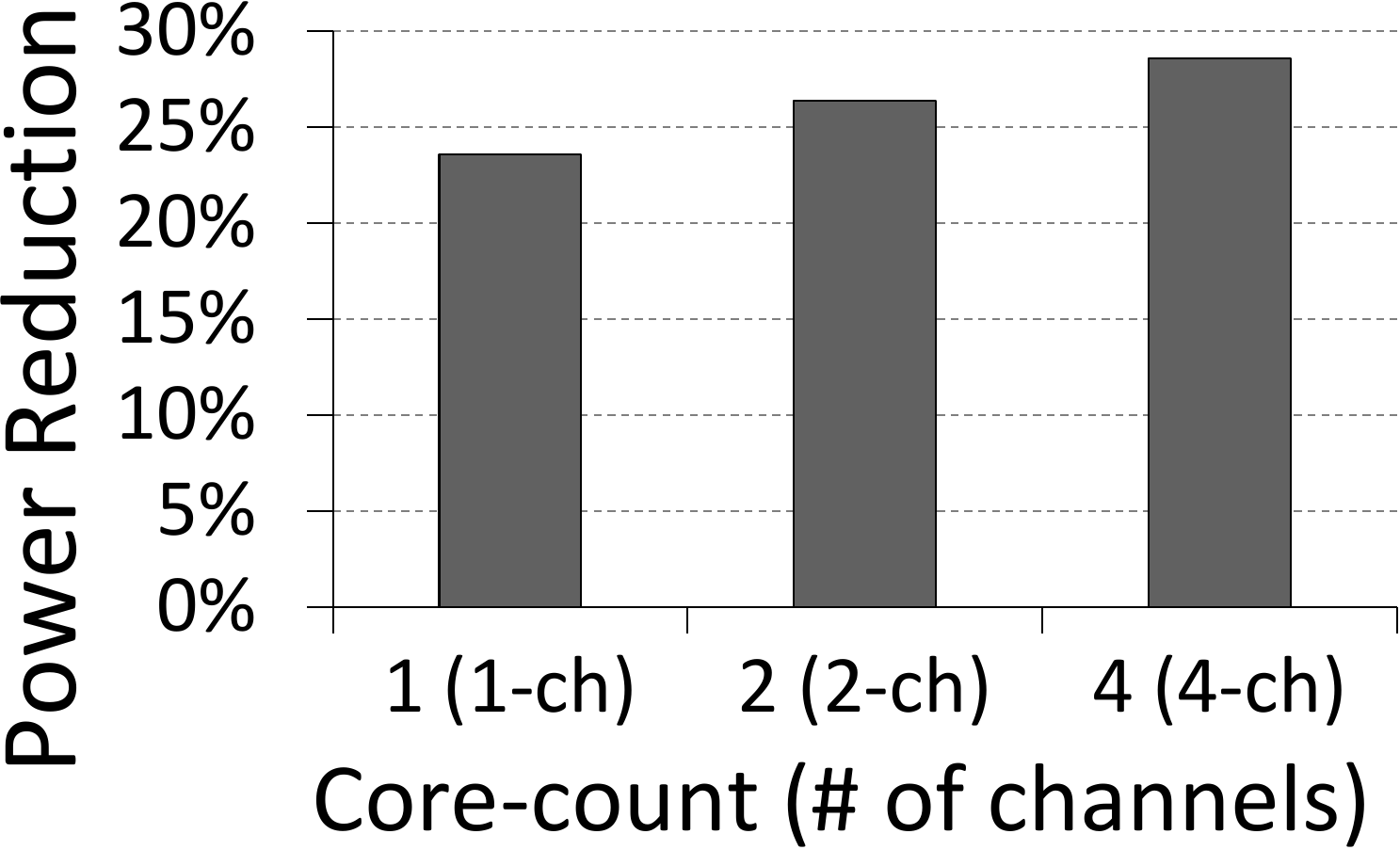}
    \caption{Power Consumption}
    \label{fig:result_power}
  \end{subfigure}
  \caption{IPC Improvement \& Power Consumption}
  \label{fig:result_cores}
\end{figure}

{\bf Sensitivity to Near Segment Capacity.} The number of rows in the near
segment presents a trade-off, since increasing the near segment's size
increases its capacity but also increases its access latency.
Figure~\ref{fig:result_single_sensitive} shows the performance improvement of our
proposed mechanisms over the baseline as we vary the near segment size.
Initially, performance improves as the number of rows in the near segment since
more data can be cached. However, increasing the number of rows in the near
segment beyond 32 reduces the performance benefit due to the increased
capacitance.

\begin{figure}[ht]
  \includegraphics[width=\linewidth]{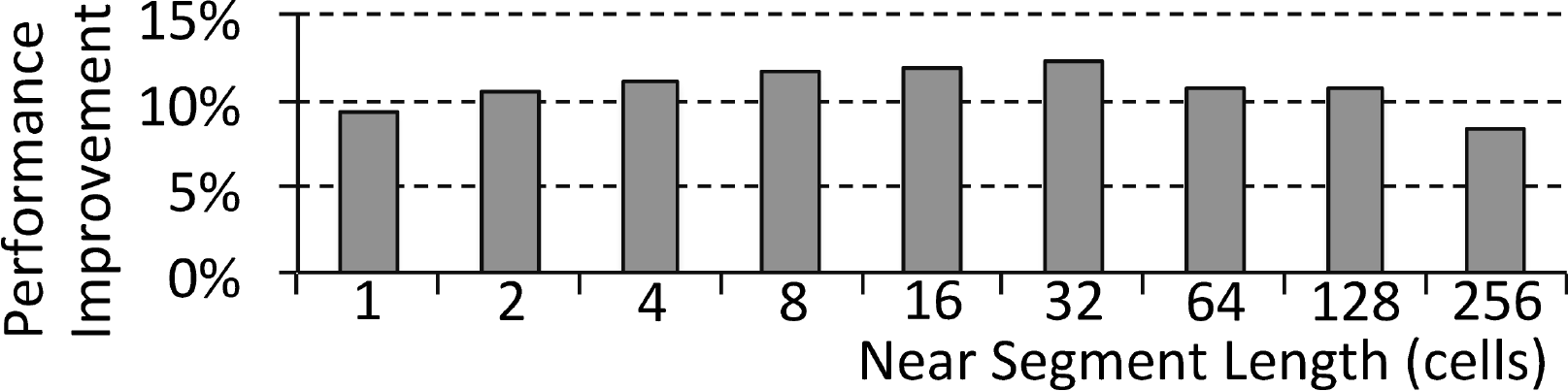}
  \caption{Effect of Varying Near Segment Capacity}
  \label{fig:result_single_sensitive}
\end{figure}

{\bf Other Results.} In our HPCA-19 paper, we provide a detailed analysis of
how timing parameters and power consumption vary when varying the near segment
length, in Section 4 and 6.3, respectively. We also provide a comprehensive
evaluation of the mechanisms we build on top of the TL-DRAM substrate for
single- and multi-core systems in Section 8.

All of our results are gathered using an in-house version of
Ramulator~\cite{kim-cal2015}, an open-source DRAM simulator~\cite{ramulator},
which is integrated into an in-house processor simulator.

\section{Significance}

\subsection{Novelty}

To our knowledge, our HPCA-19 paper is the first to enable latency
heterogeneity in DRAM without significantly increasing cost-per-bit and to
propose hardware/software mechanisms that leverage this latency heterogeneity
to improve system performance. We make the following major contributions.

{\bf A Cost-Efficient Low-Latency DRAM.} Based on the key observation that long
internal wires (bitlines) are the dominant source of DRAM latency, we propose a
new DRAM architecture called Tiered-Latency DRAM (TL-DRAM). To our knowledge
this is the first work to enable low-latency DRAM without significantly
increasing the cost-per-bit. By adding a single isolation transistor to each
bitline, we carve out a region within a DRAM chip, called the near segment,
that is fast and energy-efficient. This comes at a modest overhead of 3\%
increase in DRAM die-area. While there are two prior approaches to reduce DRAM
latency (using short bitlines~\cite{rldram, fcram}, adding an SRAM cache in
DRAM~\cite{hidaka-ieeemicro1990, cdram, esdram, cached-dram}), both of these
approaches significantly increase die-area due to additional sense-amplifiers
or additional area for SRAM cache, as we evaluate in our paper~\cite{tldram}.
Compared to these prior approaches, TL-DRAM is a much more cost-effective
architecture for achieving low latency.

\mycolor{There are many works that reduce {\em overall memory access latency}
by modifying DRAM, the DRAM-controller interface, and DRAM controllers.} These
works enable more parallelism and bandwidth~\cite{salp, dsarp, rowclone,
lee-taco2016}, reduce refresh counts~\cite{liu12, liu13, khan14,
venkatesan-hpca2006, avatar}, accelerate bulk operations~\cite{rowclone,
seshadri-cal2015, seshadri-micro2015, chang-hpca2016}, accelerate computation
in the logic layer of 3D-stacked DRAM~\cite{ahn-isca2015a, ahn-isca2015b,
zhang-hpca2014, msa3d}, enable better communication between CPU and other
devices through DRAM~\cite{lee-pact2015}, leverage process variation and
temperature dependency in DRAM~\cite{aldram}, leverage DRAM access
patterns~\cite{hassan-hpca2016}, \mycolor{reduce write-related latencies by
better designing DRAM and DRAM control policies~\cite{chatterjee-hpca2012,
lee-techreport2010, seshadri-isca2014}, and reduce overall queuing latencies in
DRAM by better scheduling memory requests~\cite{stfm-micro07, parbs, atlas,
tcm, subramanian-tpds2016, subramanian-iccd2014, rlmc, usui-taco2016}. Our
proposal is orthogonal to all of these approaches and can be applied in
conjunction with them to achieve higher latency and energy benefits.}

{\bf Inter-Segment Data Transfer.} By implementing latency heterogeneity within
a DRAM subarray, TL-DRAM enables efficient data transfer between the fast and
slow segments by utilizing the bitlines as a wide bus. This mechanism takes
advantage of the fact that both the source and destination cells share the same
bitlines. Furthermore, this inter-segment migration happens only within a DRAM
bank and does not utilize the DRAM channel, thereby allowing concurrent
accesses to other banks over the channel. This inter-segment data transfer
enables fast and efficient movement of data within DRAM, which in turn enables
efficient ways of taking advantage of latency heterogeneity.

Son et al. proposes a low latency DRAM architecture~\cite{ahn13} that has fast
(long bitline) and slow (short bitline) subarrays in DRAM. This approach
provides largest benefit when allocating latency critical data to the low
latency regions (the low latency subarrays. Therefore, overall memory system
performance is sensitive to the page placement policy. However, our
inter-segment data transfer enables efficient relocation of pages, leading to
dynamic page placement based on the latency criticality of each page.

\subsection{Potential Long-Term Impact}

{\bf Tolerating High DRAM Latency by Enabling New Layers in the Memory
Hierarchy.} Today, there is a large latency cliff between the on-chip last
level cache and off-chip DRAM, leading to a large performance fall-off when
applications start missing in the last level cache. By introducing an
additional fast layer (the near segment) within the DRAM itself, TL-DRAM
smoothens this latency cliff.

Note that many recent works added a DRAM cache or created heterogeneous main
memories~\cite{lee-isca2009, lee-ieeemicro2010, qureshi-isca2009, timber, rbla,
ramos-ics11, satish-date11, meza-weed13, hrm-dsn2014, nil-micro2012,
ren-micro2015, li-corr2015, pdram-dac09} to smooth the latency cliff between
the last level cache and a longer-latency non-volatile main memory, e.g., Phase
Change Memory~\cite{lee-isca2009, lee-ieeemicro2010,
qureshi-isca2009}, or to take advantage of the advantages of multiple
different types of memories to optimize for multiple metrics. Our approach is
similar at the high-level (i.e., to reduce the latency cliff at low cost by
taking advantage of heterogeneity) yet we introduce the new low-latency layer
within DRAM itself instead of adding a completely separate device.

{\bf Applicability to Future Memory Devices.} We show the benefits of TL-DRAM's
asymmetric latencies. Considering that most memory devices adopt a similar cell
organization (i.e., a 2-dimensional cell array and row/column bus connections),
our approach of reducing the electrical load of connecting to a bus (bitline)
to achieve low access latency can be applicable to other memory devices.

Furthermore, the idea of performing inter-segment data transfer can also
potentially be applied to other memory devices, regardless of the memory
technology. For example, we believe it is promising to examine similar
approaches for emerging memory technologies like Phase Change
Memory~\cite{lee-isca2009, qureshi-isca2009, qureshi-micro2009, meza-iccd2012,
yoon-taco2014, lee-cacm2010} or STT-MRAM~\cite{kultursay-ispass2013,
wang-islped2014}, as well as the NAND flash memory
technology~\cite{luo-msst2015, yu-hpca2015, yu-dsn2015, cai-sigmetrics2014,
cai-iccd2013}.

{\bf New Research Opportunities.} The TL-DRAM substrate creates new
opportunities by enabling mechanisms that can leverage the latency
heterogeneity offered by the substrate. We briefly describe three
directions, but we believe many new possibilities abound.

{\setlength{\leftmargini}{0.15in}
\begin{itemize}\itemsep0pt\parskip0pt\vspace{-0.1in}

	\item {\em New ways of leveraging TL-DRAM.} TL-DRAM is a substrate that can
	be utilized for many applications. Although we describe two major ways of
	leveraging TL-DRAM in our HPCA-19 paper, we believe there are several more
	ways to leverage the TL-DRAM substrate both in hardware and software. For
	instance, new mechanisms could be devised to detect data that is latency
	critical (e.g., data that causes many threads to becomes
	serialized~\cite{ebrahimi-micro2011, dm-isca10, bis, acs, uba} or data that
	belongs to threads that are more latency-sensitive~\cite{atlas,
	tcm, mise, usui-taco2016, sms, medic-pact, subramanian-iccd2014,
	subramanian-tpds2016, subramanian-micro2015}) or could become latency
	critical in the near future and allocate/prefetch such data into the near
	segment.

	\item {\em Opening up new design spaces with multiple tiers.} TL-DRAM can be
	easily extended to have multiple latency tiers by adding more isolation
	transistors to the bitlines, providing more latency asymmetry. (Our HPCA-19
	paper provides an analysis of the latency of a TL-DRAM design with three
	tiers, showing the spread in latency for three tiers.) This enables new
	mechanisms both in hardware and software that can allocate data appropriately
	to different tiers based on their access characteristics such as locality,
	criticality, etc.

	\item {\em Inspiring new ways of architecting latency heterogeneity within
	DRAM.} To our knowledge, TL-DRAM is the first to enable latency heterogeneity
	within DRAM by significantly modifying the existing DRAM architecture. We
	believe that this could inspire research on other possible ways of
	architecting latency heterogeneity within DRAM or other memory devices.

\end{itemize}
}

\bibliographystyle{abbrv}
\setstretch{0.20}
\begin{footnotesize}
\bibliography{paper}
\end{footnotesize}

\end{document}